\documentclass{aa}
\usepackage{astro_bib_macro}
\usepackage{natbib}
\bibpunct{(}{)}{;}{a}{}{,}

\usepackage{epsf}

\begin{document}

\newcommand{\iras}{} \def\iras/{G29.96}
\newcommand{\mysp}{}\def\mysp/{}
\newcommand{\allo}[3]{\ion{#1\mysp/}{#2}\,#3$\mu$m}
\newcommand{\forb}[3]{[\ion{#1\mysp/}{#2}]\,#3$\mu$m}
\newcommand{\sforb}[3]{[\ion{#1\mysp/}{#2}]\,#3$\mu$m}
\newcommand{\dforb}[4]{[\ion{#1\mysp/}{#2}]\,#3,\,#4$\mu$m}
\newcommand{\rforb}[4]{[\ion{#1\mysp/}{#2}]\,#3/\,#4$\mu$m}
\newcommand{\ro}{} \def\ro/{r$_{[\ion{O\mysp/}{iii}]}$}
\newcommand{\hi}{}\def\hi/{\ion{H\mysp/}{i}}
\newcommand{\hii}{}\def\hii/{\ion{H\mysp/}{ii}}
\newcommand{\uchii}{}\def\uchii/{UCHII}
\newcommand{\bra}{}\def\bra/{Br$\alpha$}
\newcommand{\masection}{}\def\masection/{Sect.}

\newcommand{\papit}{}\def\papit/{Paper~I}
\newcommand{\papip}{}\def\papip/{(Paper~I)}
\newcommand{\papiit}{}\def\papiit/{Paper~II}
\newcommand{\papiip}{}\def\papiip/{(Paper~II)}

\def\teff{\ifmmode T_{\rm eff} \else $T_{\mathrm{eff}}$\fi}
\def\lg{$\log g$}
\def\msun{\ifmmode M_{\odot} \else M$_{\odot}$\fi}
\def\zsun{\ifmmode Z_{\odot} \else Z$_{\odot}$\fi}
\def\lsun{\ifmmode L_{\odot} \else L$_{\odot}$\fi}

\title{A photoionization model of the compact \hii/ region
G29.96-0.02\thanks{Based on observations with ISO, an ESA project with
instruments funded by ESA Member States (especially the PI countries:
France, Germany, the Netherlands and the United Kingdom) and with the
participation of ISAS and NASA.}
       }

\titlerunning{A photoionization model of \iras/-0.02}
\authorrunning{C. Morisset et al.}

\author{C. Morisset \inst{1}
    \and D.~Schaerer \inst{2}
    \and N.~L.~Mart\'{\i}n-Hern\'andez \inst{3}
    \and E.~Peeters \inst{3,4}
    \and F.~Damour \inst{1}
    \and J.-P.~Baluteau \inst{1}
    \and P.~Cox \inst{5}
    \and P.~Roelfsema \inst{4}
        }

\offprints{C. Morisset}
\mail{Christophe.Morisset@astrsp-mrs.fr}

\institute{Laboratoire d'Astrophysique de Marseille, CNRS, BP 8,
           F-13376 Marseille Cedex 12, France
      \and
       Laboratoire d'Astrophysique, Observatoire Midi-Pyrénées, 14,
           Av. E. Belin, F-31400 Toulouse, France
      \and
       Kapten Astronomical Institute, P.O. Box 800, NL-9700 AV
           Groningen, The Netherlands
      \and
       SRON, P.O. Box 800, NL-9700 AV
           Groningen, The Netherlands
      \and
       Institut d'Astrophysique Spatiale, Bat. 121, Universit\'e de
           Paris XI, F-91405 Orsay France
           }

\date{}


\abstract{
We present a detailed photoionization model of G29.96-0.02
(hereafter G29.96), one of the brightest Galactic Ultra
Compact \ion{H}{ii} (UCHII) regions in the Galaxy. This source has
been observed extensively at radio and infrared
wavelengths. The most recent data include a complete ISO (SWS and LWS)
spectrum, which displays a remarkable richness in atomic
fine-structure lines. The number of observables is
twice as much as the number available in previous studies. In
addition, most atomic species are now observed in two ionization 
stages. 
The radio and infrared data on G29.96 are best reproduced using a
nebular model with two density components: a diffuse (n$_e\sim 680$
cm$^{-3}$) extended ($\sim$1 pc) component surrounding a compact
($\sim$0.1 pc) dense (n$_e\sim 57000$ cm$^{-3}$) core.  The
properties of the ionizing star were derived using state-of-the-art
stellar atmosphere models. {\em CoStar} models yield an effective
temperature of $\sim 30^{+2}_{-1}$ kK whereas more recent non-LTE 
line blanketed atmospheres with stellar winds indicate
somewhat higher values, \teff\ $\sim$ 32--38 kK. This range in \teff\ is
compatible with all observational constraints, including
near-infrared photometry and bolometric luminosity. 
The range 33-36~kK is also compatible with the
spectral type O5-O8 determined by \citet{WH97} when
recent downward revisions of the effective temperature scale
of O stars are taken into account. The age of the ionizing
star of G29.96 is found to be a few $10^6$~yr, much older than
the expected lifetime of UCHII regions. Accurate gas phase
abundances are derived with the most robust results being 
Ne/S~=~7.5 and N/O~=~0.43 (1.3 and 3.5 times the solar
values, respectively).
 }

\maketitle

            \section{Introduction}
\label{sec:intro}

The properties of young massive stars are still
poorly known. This is due to the fact that massive stars
are relatively rare and that their lifetime is short. 
In their early stages, massive stars are still embedded 
in their parental molecular clouds and suffer heavy 
extinction. Our knowledge of the stellar energy 
distribution in the visible and the ultraviolet of 
young massive stars therefore relies on indirect 
measurements, such as atomic fine-structure lines observed at 
infrared wavelengths. A comparison of the line 
fluxes with predictions of detailed photo-ionizing 
codes coupled with stellar atmosphere models allows 
one to constrain the properties of the ionizing stars.

Recently, \citet{PaperI} and \citet{PaperII}, hereafter 
Papers I \& II, presented
the results of an infrared spectral survey of 34 galactic 
compact \hii/ regions based on complete ISO grating spectra. 
For most of the sources, the ISO data display a remarkable 
richness in  spectral lines including all the fine-structure
lines of N, O, ne, S, and Ar. Using additional data from 
the literature, the elemental abundances and their variation
across the Galactic disc were derived (Paper~II). 

Some of the \hii/ regions in the sample are well known and were studied
in detail at other wavelengths. Those additional data provide useful
contraints to derive the properties of the ionizing star and to fine
tune the elemental abundance estimates. One such source is the compact
\hii/ region IRAS 18434-0242 (G29.96-0.02, herafter G29.96). This
compact source has one of the richest ISO spectrum of the entire sample
(Paper~I) and is one of the few compact \hii/ regions for which the
ionizing star has been identified and characterized through direct
infrared spectroscopy \citep{WH97,HLR02}.

The availability of high-quality data on \iras/ 
together with the recent developments of the models of
stellar atmosphere of massive stars prompted us to make
a detailed photionization model of this source with the
aim to further constrain the nature of the ionizing stars
and to derive accurate elemental abundances. The paper
is organized as follows:
the observational properties are summarized
in Sect.\ 2; Sect.\ 3 describes the photoionization code, the 
input stellar parameters and the methodology; Sect.\ 4 
presents the results of the best model which are discussed
in Sect.\ 5; the conclusions are given in Sect.\ 6. 
 
            \section{G29.96-0.02: Observational facts}
        \label{sec:iras}

The compact \hii/ region \iras/ is one of the brightest radio and
infrared sources in our Galaxy. Its morphology is a classical
example of a cometary-like compact \hii/ region \citep{WC89}
in interaction with a molecular cloud \citep[e.g.,][hereafter
PMB99]{PMB99}.
\iras/ has been studied in detail at
infrared and radio wavelengths and the following summarizes the main
results together with the essential properties of this compact \hii/
region.

\subsection{Observations of fine-structure lines}
\label{sub:isoobs}
\label{sub:otherobs}
\label{sub:kao}

The fine-structure lines observed by the SWS and LWS
spectrometers are tabulated in \papit/ where their
observed fluxes associated error bars are given as well as the entire
ISO spectrum of this source. In addition, 11 hydrogen recombination lines were
detected in \iras/ 
and were used to determine the interstellar extinction \papiip/. We will use
\bra/ in the modeling process (see Sect.~\ref{sub:conv} and
Table~\ref{tab:model}) as this is the brightest \hi/ line and
one of the least affected by extinction.
The ISO lines corrected from the interstellar extinction
(using A$_K$=~1.6 and the extinction law derived in \papiit/) are
given in Column~2 of Table~\ref{tab:results}.

Infrared fine-structure lines have previously been observed with KAO by
\citet{HHF81}, who  measured the lines \forb{Ar}{ii}{6.98},
\forb{Ar}{iii}{8.98}, \forb{S}{iii}{18.7} and \forb{Ne}{ii}{12.8},
in agreement with the ISO values within 20~\%.

Similarly, \citet{screh95} reported observations
of \forb{S}{iii}{33.6}, \forb{Ne}{iii}{36.0}, \forb{O}{iii}{51.8},
\forb{N}{iii}{57.3} and \forb{O}{iii}{88.3} differing from the ISO
values by less than 10~\%, except the
\forb{S}{iii}{33.6} and the \forb{O}{iii}{88.3} fluxes which are both
$\sim 30$~\%
higher in our sample. The difference of aperture sizes and pointing
between KAO and ISO can partly be responsible of these discrepancies. 

Maps of \forb{Ne}{ii}{12.8} were obtained by \citet{LBG82} and
\citet{WMTM98}. These authors found an integrated
flux of 84~\% and 64~\%
of our value, over a size of 10''$\times$10'' and a diameter of 30''
respectively.

\subsection{Radio observations: core/halo structure and constraints
  on the He ionization structure}
\label{sub:corehalo}
\label{sub:morph}

\iras/ has been observed at radio frequencies using various spatial
resolutions. 
At 2~cm the observed flux densities range from 2.7 to 4.6~Jy
\citep{WC89,ACHK94,FGCV95,WCSHC97}, at 6~cm from 1.4 to 3.6~Jy
\citep{WC89,ACHK94}, and at 21~cm from 0.9 to
2.6~Jy \citep{CH95,Kim01}. As usually the case for \uchii/ regions, the
source presents a diffuse 
emission. Therefore the highest spatial resolution observations may
miss part of the radio flux density. We have adopted the following values: 3.9,
3.4 and 2.6~Jy at 2, 6, and 21~cm respectively, favoring the highest
values to take into account the diffuse emission.
The resulting number of Lyman continuum photons 
($1.8 \times 10^{49}$, cf.\ below ) 
is higher than the value of $4 \times 10^{48}$
derived in \papiit/, but this last value was obtained for a uniform
electron density of 10$^4$ cm$^{-3}$ and a size of 7 arcsec.

\iras/ is characterized by a strong
edge-brightened core/"head", with a low surface-brightness "tail"
of emission trailing off opposite the bright edge. Due to this
extended emission, part of \iras/ cannot be strictly called Ultra
Compact. 
\citet{WC89} obtained n$_e$ = 8.5$\times 10^4$ cm$^{-3}$ in the arc
and a n$_e$ 5-10 
times lower in the tail of the nebula, while \citet{ACHK94}
estimate n$_e$ = 5.$\times 10^4$ cm$^{-3}$
in the leading arc and n$_e$ = 2.$\times 10^4$ cm$^{-3}$ in the tail.
\citet{screh95} obtained $\rm n_e = 1500\ cm^{-3}$ from the
\rforb{O}{iii}{51.8}{88.3} lines ratio.
From the set of the ISO observables available for \iras/, 3 line ratios
can be used as density
diagnostics: \rforb{O}{iii}{51.8}{88.3}, \rforb{S}{iii}{18.7}{33.6} and
\rforb{Ne}{iii}{15.5}{36.0}. Unfortunately, the two last ratios suffer
large calibration uncertainties (25 \% at 1 $\sigma$ error, see \papit/) 
and therefore only
\rforb{O}{iii}{51.8}{88.3} (hereafter \ro/) can be safely used to
derive the gas density. From \ro/ = 2.4 an
electron density n$_e$ $\sim$ 800 cm$^{-3}$ is derived \papiip/.

This apparent discrepancy between the densities determined from
fine-structure lines of oxygen 
and from radio observations have already been pointed
out by \citet{ACW97} who suggested a core/halo description of \iras/. 
Faint diffuse halos are commonly observed in
the radio continuum maps of \uchii/ regions
\citep[e.g.,][]{GRM93,FGCV95,ACA96,KWH99}. Recently, \citet{Kim01}
found extended emission at 21~cm linked to the bright spot of \iras/.
\bra/ images \citep[PMB99]{LH96,WCSHC97,WH97} also support this
morphology.

The observational evidence of a core/halo morphology 
has led us to model G29.96 with two components as
outlined in \masection/ \ref{sub:2comp}.

\citet{Kim01} have observed the He76$\alpha$ and H radio recombination lines 
in \iras/. For various positions, including the region considered here,
they obtain a He$^+$ abundance of $\sim$ 0.068--0.076. 
Assuming a normal helium abundance of 0.1 this implies that He is
predominantly singly ionized helium in \iras/.
This provides a useful constraint on the temperature of the ionizing
source (see Sect.\ \ref{sub:otherstars}).

\subsection{Heliocentric and galactocentric distances}
\label{sub:dist}

With an average global radio recombination lines LSR velocity of 95
km/s \citep{ACHK94}, adopting a galactocentric
radius of the Sun of 8.5 kpc, a kinematic heliocentric distance for
\iras/ between 5.5 and 9.5~kpc 
is derived using a standard galactic
rotation curve with a rotation speed of 220 km/s at the Suns position.
Previous investigators assumed the average between
the near and far distances.

The extinction along the line of sight to \iras/ was studied by PMB99 
using galactic
\hi/ and CO surveys. They found A$_K\sim$1 at a distance of 5 kpc, and
A$_K\sim$3 at a position corresponding to the tangent point (at about 7.5
kpc). These findings provide a crucial argument to consider that the
near distance should be more appropriate. Hereafter,
we will adopt an heliocentric
distance of 6 kpc (+1.0,-0.5)  for \iras/\footnote { At 6.0 kpc,
1$^{\prime\prime}$ = 0.029 pc = 8.95$\times 10^{16}$ cm}, implying
R$_G$=4.5 kpc (+/-0.3). 

\subsection{Ionizing star(s)}
\label{sub:stars}

A detailed search for stars embedded in the \hii/ region and the adjacent
molecular hot core has been performed in the near-infrared
\citep[PMB99]{LH96,WCSHC97,WH97}.
\citet{WCSHC97} and PMB99 have in particular revealed the existence
of a cluster of about 18 OB-type stars, or their progenitors,
embedded in the cloud.
The same authors have convincingly identified the bright star at the
center of the arc of radio emission as the exciting star, or at least as
the primary source of ionization. There is no evidence for an infrared
excess in the K band, suggesting that any remaining disk should be
optically thin, and therefore that the star is no longer accreting
mass.
From near infrared spectroscopic observations, \citet{WH97} were able 
to constrain the spectral type of the ionizing source. 
A spectral type between O5 and O8 (and no constraint on the luminosity
class) was found. In contrast, a recent study by 
\citet{KBHC01} reports a spectral type as early as O3.
From their spectroscopic and photometric data \citet{WH97} and
\citet{WCSHC97} already deduced an 
evolutionary age of about 1-2$\times 10^6$ years for a single or
binary star, in apparent contradiction with the estimated age of the
\uchii/ region ($\sim 10^5$ yr). 

The overall SED of \iras/ derived by \citet{WCSHC97},
the near-IR photometry, and spectral types provide important
constraints on the fundamental properties (\teff, luminosity,
age) of the ionizing source. A detailed discussion is given
in Sect.\ \ref{sub:otherstars}.

            \section{Photoionization modeling}
\label{sec:model}

\subsection{Photoionization code}
\label{sub:nebu}

The models are performed using the detailed
photoionization code NEBU \citep{MP96,P02} which consistently computes the
line fluxes without 
any hypothesis on the ionization structure of the gas, especially
without using ionization correction factors ({\em icf}'s). However, it
does require assumptions about the geometry, density, and 
pressure structure of the nebula.

The computation is performed in a spherical geometry
and at each radius from the central ionizing source, the electron
temperature, the electron and ions densities, and the line emissivities
are determined solving the ionic and thermal equilibrium equations.
The inputs for the model are the description of the ionizing
flux, using e.g. an effective temperature and a luminosity (see \masection/
\ref{sub:costar}), and the gas distribution
(assuming e.g. constant pressure through the shell) with a set of abundances.

The elements taken into account and for which lines fluxes are
predicted are: H, He, C, N, O, Ne, Mg, Si, S, Cl, Ar, Fe, Ca and Ni.
Self-absorption effects of the radio continuum are computed in a
spherical geometry approximation. 

Absorption of incoming and diffuse photons by
dust is considered, the number ratio of dust grains over the
number of hydrogen atoms being a parameter of the model. The optical
properties of dust grains were made available to us by Ryszard
Szczerba (private communication). The adopted dust composition is 50~\%
"astronomical" silicates and 50~\%
graphite \citep{DL93}. No quantum heating by very small grains are
taken into account in the version of NEBU used for this work.

\begin{figure*}
\epsfxsize=18.cm  \epsfbox{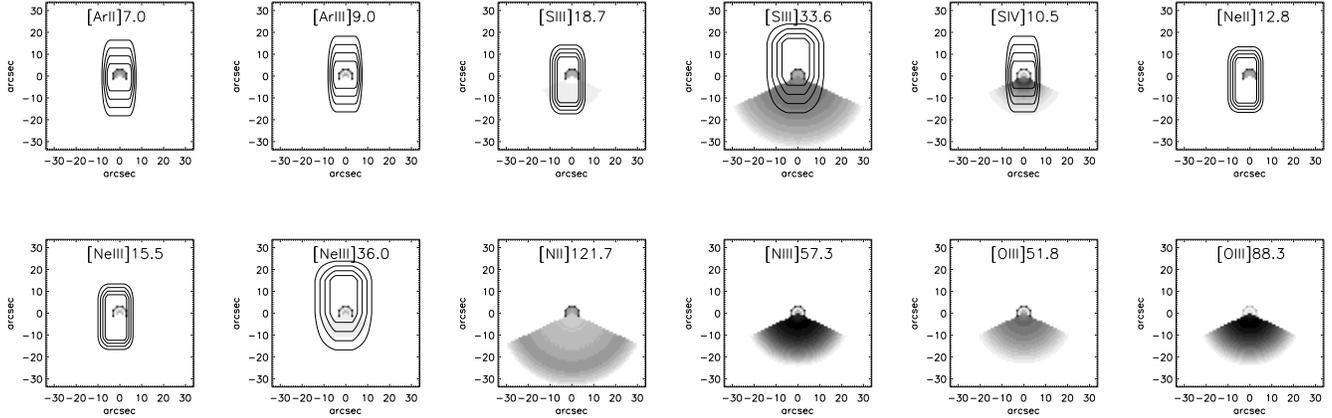}
\caption{Emission lines maps of the best model, in grey linear
scale. The two componets are presented with angular size related to
their corresponding covering factors.
Contours for the ISO SWS beam profile are superposed, for transmissions of 20,
40, 60, 80~\% 
of the flux. In the four latest maps, the LWS aperture size is of
order of the images, and not 
shown. Note that the SWS aperture is not always centered. 
\label{fig:maps}}
\end{figure*}

\subsection{Atmosphere models}
\label{sub:costar}
The ionizing photons distribution is taken from the recent {\em CoStar} 
atmosphere models, which include non-LTE effects, line blanketing, and
stellar winds \citep{SC97}.
The {\em CoStar} atmospheres have been compared to earlier
predictions (Kurucz models and plane parallel non-LTE atmosphere
models). Their implications on nebular studies have also
been discussed by \citet{Sch97} and \citet{SS99}.

The predicted ionizing fluxes have been ``tested'' by the following
indirect means:
\begin{itemize}
\item[{\em 1)}] the {\em CoStar} SEDs reproduce well the ionic
Ne$^{++}$/O$^{++}$ 
abundance ratios in Galactic \hii/ regions observed from optical and
IR data, thereby solving the so-called [Ne~{\sc iii}]
problem \citep{Sch97}\footnote{The atmosphere models calculations of
\citet{SYPR96} have not been confirmed by the Munich group. The 
spectra from their latest grid of atmosphere models including non-LTE 
line blocking and blanketing \citep{PHL01} predict 
[Ne~{\sc iii}]/ [Ne~{\sc ii}] ratios in rough, though somewhat marginal 
agreement with observations (\citet{GSL02}, cf.\ Sect.\ 
\ref{sub:comp_star}).}.
This supports the important increase of the ionizing spectra
predicted at energies $\ga$ 35--40~eV.

\item[{\em 2)}] the observed increase of the ratio of 
the He$^0$ over H$^0$ 
ionizing photons between stellar effective temperatures of $\sim$ 30000 
and 40000 K is well reproduced \citep{S00}.
This quantity, sensitive to the ionizing flux above 13.6 and 24.6~eV,
is constrained by optical He~{\sc i} recombination line 
measurements in the sample of \citet{KBFM00} of
\hii/ regions  with known stellar content.
\item[{\em 3)}] Tailored photoionization models for two nebulae with
exciting stars of spectral types O3-O4 and O7 respectively have been
constructed by \citet{ODSS00} using {\em CoStar} spectra.
While overall the spectra are found to yield a good agreement with the
observed nebular line ratios for all objects (confirming also the
first point), there 
is some indication of an overprediction of the spectrum above $\sim$ 
35--40~eV for the early-type object (DEM L323). 
We note, however, that for DEM L243, this discrepancy is not found if a
cooler spectrum is adopted (approximately model B2 instead of C2 preferred
by Oey et al.), as would be expected for the region whose ionizing 
flux is likely dominated by the O7If supergiant instead of the O7V((f)) 
dwarf \citep[cf.][]{VGS96}.

\end{itemize}

For the range of stellar temperatures of interest here ($T_{\rm eff}
\sim$ 28--34 
kK; cf.\ below) only 1 and 2 are directly applicable here.
Indeed, at these relatively cool effective temperatures the uncertainties in
the {\em CoStar} models are expected to be the largest, 
as stressed by \citet{SC97}.
Although the constraints on the atmosphere models are still rather limited,
we conclude from the above comparisons that 
the adopted
model atmospheres should provide a reasonable description of the ionizing
fluxes. Comparisons with other model atmospheres are discussed
in Sect.\ \ref{sub:atmomod}.

From the 27 {\em CoStar} models available on the Web \citep{SC97}, a 
finer grid of spectral energy distributions was constructed.
For any effective temperature \teff\ and luminosity
$L$ in the range covered by the {\em CoStar} models, an interpolation is
performed between the four nearest models of this grid, using the
square inverse of the distance in the $\log \teff$--$\log L$ plane to
determine the weights of the four {\em CoStar} models. 
The four {\em CoStar} models are first divided by the corresponding
blackbody spectra, then averaged using the weights previously
determined and the result is finally multiplied by the blackbody
spectrum of the desired \teff\ and $L$, avoiding the {\em diktat} of
the most intense {\em CoStar} model.

\subsection{Two-component model}
\label{sub:2comp}

In order to reproduce both \ro/ and the radio flux densities  which
imply a higher density (see
\masection/~\ref{sub:corehalo}), a two-component model is used to describe
\iras/. The two components differ by their densities; the lower and
higher density components are named component~1 and 2 respectively.

The present model consists of a simple linear combination of two
independent runs of NEBU in a spherical case and under isobar
approximation. Both components are assumed to be radiation
bounded. The gas is supposed to be homogeneously distributed in
each component, with an inner radius to reproduce an empty
cavity. The coefficient applied to each component to obtain the
fluxes of the emission lines is the {\em covering factor},
i.e. the angular size over 4 $\pi$ of each component as seen by
the central source. The sum of the two covering factors is set to
one, i.e. we do not consider that photons could escape from the
\hii/ region. In this model, there is no diffuse
radiation exchanged between component~1 and 2, and no effects of
shielding of one 
component by the other is taken into account as: component~2 has a
very small geometrical thickness compared to component~1, and
component~1 is very optically thin at radio wavelengths. A
two-component model must be seen as an approximation, describing
the two first moments of a certainly more complicated
gas distribution.

The most important effect of this 2-density medium is in the
localization of the lines emission, which depends on their
critical densities. For example, the [\ion{O}{III}] and [\ion{N}{III}]
lines are 
emitted only by the lower density part of \iras/, because these lines
are collisionally de-excited in the densest region (see
\papiit/ for the critical densities of all the lines).

\subsection{The ISO beam profiles}

Component~1 of the nebula will have a larger geometric
extension than component~2 due to its lower density. 
The beam sizes of the ISO instruments, which vary from
14-33 arcsec for the SWS to $\sim$~80 arcsec for the LWS, have a 
drastic impact when comparing/dividing different emission
lines.
The beam profiles, from \citet{G99}, are used to apply a
correction to the predicted line fluxes.
For each component and line, an intensity map is generated by
projection of the line emissivity on a sky plane. An ``ISO'' mask is
then applied, according to the detector beam corresponding to the line
wavelength.

Fig.~\ref{fig:maps} shows the intensity maps obtained for our best
model (presented in \masection/~\ref{sec:res}) for 8 atomic
lines. Contours of the
corresponding ISO SWS apertures are superimposed on the image. For
the four last images corresponding to lines observed by LWS, the aperture size
is larger than the image.
For most of the lines, only the core is visible with the
adopted linear intensity scale. Note that, despite the absence of
visible extended 
emission, for most of the lines the contribution of the extended part
is about that of the core.
The low density extended component is clearly seen for the low
critical density lines for which the contribution of the core is very
low (the [\ion{N}{II}], [\ion{N}{III}],
[\ion{O}{III}] and [\ion{S}{III}] lines). 
The effect of neglecting the finite aperture size is clearly
illustrated. Note also that the profile of the SWS aperture, as described
by \citet{G99}, is not symmetrical. We did not attempt to exactly adjust the
orientation of these profiles to the observations of \iras/, as the effect of
the asymmetry in the profiles is negligible compared to
the effect of departure from spherical symmetry of the object itself.

        \section{Results}
    \label{sec:res}

        \subsection{Convergence procedure}
\label{sub:conv}

Table~\ref{tab:model} lists the parameters of the best two-component
model and the observable used to constrain each 
parameter. In the following we describe the convergence procedure of
this model.

The [\ion{O}{iii}] lines are collisionaly de-excited in component~2
and trace mainly component~1, while the other density diagnostic lines
are emitted by both components of the nebula.
Therefore the \rforb{O}{iii}{51.8}{88.3} ratio is used to constrain the
density of component~1. The density of component~2 is constrained by
the 6~cm radio flux density. The ratio of the two covering factors 
is determined by fitting the \bra/ line flux, the sum beeing 
fixed to 1.

\begin{figure}
\epsfxsize=8.cm  \epsfysize=7.5cm \epsfbox{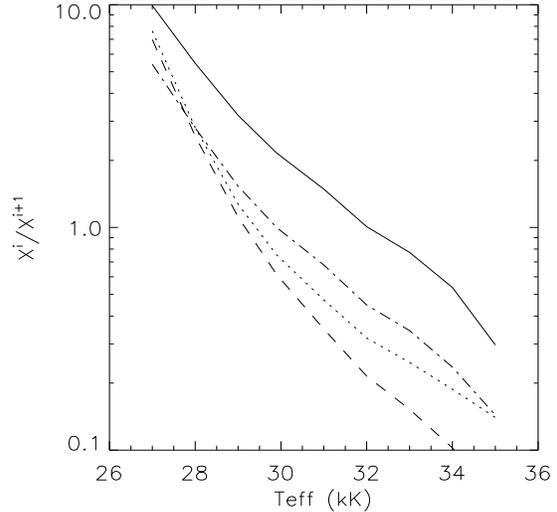}
\caption{Variation of the four [X$^i$]/[X$^{i+1}$]
ratios with the effective temperature. The ratios correspond to: Ar
(solid) , S (dot), Ne (dashed) and N (dot-dashed) and have been
divided by their 
corresponding observed values. \label{fig:t4}}
\end{figure}

Fig.~\ref{fig:t4} shows the four available [X$^i$]/[X$^{i+1}$] line
ratios (divided by the corresponding observed values) versus 
the effective temperature of the ionizing star, namely
\forb{Ar}{ii}{6.98}/\forb{Ar}{iii}{8.98},
\forb{S}{iii}{18.7}/\forb{S}{iv}{10.5},
\forb{Ne}{ii}{12.8}/\forb{Ne}{iii}{15.5} and
\forb{N}{ii}{121.7}/\forb{N}{iii}{57.3}.
The number of ionizing photons is kept constant by adjusting
the number of ionizing stars with fixed luminosity, while the
effective temperature is 
varied. The models used for this plot are all 2-component models.
In the range of $T_{\rm eff}$ considered here (27 to 35 kK) the sensitivity of
those ratios to the effective temperature is very high 
\citep[see also ][and \papiit/]{Sch97}. 
Within a range of $\sim$ 2500 K, the four ratios provide the same
constraint on the effective temperature. However, the
\forb{Ar}{ii}{6.98}/\forb{Ar}{iii}{8.98} ratio gives a higher
effective temperature than the three other diagnostics (see \masection/
\ref{sub:atmomod}). We will 
therefore use the 
\forb{N}{ii}{121.7}/\forb{N}{iii}{57.3} ratio to determine the
effective temperature, as it is: 1) independent of the ISO beam size since
both lines were observed with LWS, and 2) emitted only by component~1,
since it is collisionaly de-excited in component~2. An effective
temperature of $\sim 30^{+2}_{-1}$ kK is then derived.

\begin{table*}
\begin{center}
\caption{Parameters of the best model and derived gas
properties. For some parameters, the observable used for the main
constraint is given. 
    \label{tab:model}
     }
    \begin{tabular}{lll}
    \hline
      \noalign{\smallskip}
    \multicolumn{3}{c}{Parameters common for both components} \cr
      \noalign{\smallskip}
    \hline

Effective temp (kK)         & 29.7      & from \forb{N}{ii}{121.7}/\forb{N}{iii}{57.3} \cr
Luminosity in log(L/\lsun)    & 5.86$^1$      & from borderline
in {\em CoStar} models grid (see \masection/~\ref{sub:comp_star}) \cr
Number of stars             & 1.5       & from 2~cm flux density (see \masection/~\ref{sub:nstars})\cr
Inner radius ($10^{17}$ cm)     & 3.        & from imaging \cr
Dust/gas ratio                  & $10^{-5}$ & from ISO continuum emission      \cr
[He]/[H] (in numbers)           & \multicolumn{2}{l}{0.1} \cr
[C]/[H]     & 1.00$\times 10^{-4}$ \cr
[N]/[H]     & 1.97$\times 10^{-4}$ & from \forb{N}{ii}{121.8} + \forb{N}{iii}{57.3} \cr
[O]/[H]     & 4.55$\times 10^{-4}$ & from \forb{O}{iii}{51.8 + 88.3} \cr
[Ne]/[H]    & 1.70$\times 10^{-4}$ & from \forb{Ne}{ii}{12.8} + \forb{Ne}{iii}{15.5}\cr
[S]/[H]     & 2.25$\times 10^{-5}$ & from \forb{S}{iii}{18.7} + \forb{S}{iv}{10.5}\cr
[Ar]/[H]    & 5.00$\times 10^{-6}$ & from \forb{Ar}{ii}{6.98} + \forb{Ar}{iii}{9.0} \cr

    \hline
      \noalign{\smallskip}
    \multicolumn{1}{r}{Parameters for:}&    Component 1 &Component 2 \cr
       \noalign{\smallskip}
   \hline
Inner n$_H$ (cm$^{-3}$) &   640. from \ro/        &   52000. from 6~cm
flux density\cr
Covering factor     &   36~\% \hspace{.4cm} and   &   64~\%  from \bra/  \cr
Filling factor          &       1.00    &       1.00    \cr
    \hline
      \noalign{\smallskip}
    \multicolumn{1}{r}{Properties for:}    & Component 1   &Component 2 \cr
      \noalign{\smallskip}
        \hline
Mean n$_e$ (cm$^{-3}$)           &               680.   &   57000.  \cr
Constant pressure (cgs)      &9.7$\times 10^{-10}$&1.1$\times 10^{-7}$\cr
Thickness (10$^{17}$ cm)&   27.2    &   0.15 \cr
Mass (M$_{\odot}$) &   29.    &   0.73  \cr
Mean T$_\hii/$ (K) & 5520 & 7230 \cr
inner U    & 8.8$\times 10^{-1}$ & 5.7$\times 10^{-3}$\cr
        \hline
    \end{tabular}
\end{center}
$^1$ {\footnotesize This luminosity multiplied by the number of stars
leads to $1.8 \times 10^{49}$ ionizing photons (N$_{\rm Lyc}$).}

\end{table*}

The total luminosity is the product of the {\em CoStar} model luminosity by the
number of stars, and the number of ionizing photons
is adjusted to reproduce the radio flux density at 2~cm. A range of stars with
different spectral types is
presently not considered.
With the available {\em CoStar} models, there is an upper limit for the star
luminosity beyond which no more models are
available and which corresponds to the 
post main sequence or Wolf-Rayet star regime. The
degeneracy of the {\em CoStar} model luminosity by the number of stars
is discussed in 
\masection/~\ref{sub:nstars}.

Once the above set of parameters is fixed, the abundance of
each species is derived by fitting their lines fluxes. As
most of the parameters have feedback effects, an iterative process
must be applied. 

To summarize, we have 17 free parameters (listed in
Table~\ref{tab:model}), 
namely: effective temperature and
luminosity of one star, number of stars, densities of both components,
ratio of the covering factors, inner radius of the empty cavity,
dust to gas ratio, filling factors of both components, and 7 abundances.
On the other hand, we have 16 observables plus the morphology of the
source as given by the radio maps.

Some parameters (the inner radius, the
dust/gas ratio, the filling factors) cannot be precisely
constrained and are set to a reasonable value. Change in their values are
discussed in \masection/~\ref{sec:disc}.
The abundance of helium and carbon does not have any
effect on the results of the model, if remaining within reasonable
values.

We finally stay with 11 free parameters to be
constrained by 17 observables. In other words, there are 6 observables
that are not used to build the model and are therefore entirely
predicted, namely: the three [X$^i$]/[X$^{i+1}$] ratios for Ne, Ar and
S, the two \forb{S}{iii}{33.6} and
\forb{Ne}{iii}{36.0} line fluxes (not used because of their large
calibration error, see \papit/ and \papiit/), and the 21~cm continuum
flux density. 
        \subsection{Results and predictions}
\label{sub:results}

Table~\ref{tab:results} lists the observations of the infrared
emission lines and the radio continuum flux densities together with
the results of the best model. The contributions of the two density
components are given separately. Most of the predicted lines fluxes and
radio flux densities agree with the observations to within the
uncertainties of the measurements. The few predictions which are off by a
larger factor are the results of well understood factors which will be 
explained hereafter.  

The three [X$^i$]/[X$^{i+1}$] ratios not used to
determine the effective temperature agree within $\sim$ 1~kK
with the \forb{N}{ii}{121.7}/\forb{N}{iii}{57.3} ratio
(see Fig.~\ref{fig:t4}). The \forb{S}{IV}{10.5} line
falls in the absorption feature due to silicate. This has been taken
into account when using the attenuation law described in
\papiit/ and
the model prediction is in very good agreement with the observation.
For \forb{Ne}{III}{15.5}, we can suspect an overprediction of the line
flux due to the used of the {\em CoStar} models, as discussed by
\citet{ODSS00} and in Sect.\ \ref{sub:costar} (see also \masection/
\ref{sub:atmomod}). 

Note that the predicted 21~cm flux density is lower than the value
observed by \citet{Kim01}, who reported complex extended radio emission
toward \iras/. Part of this diffuse emission could be due to gas
ionized by members of the stellar cluster other than the main ionizing
star of \iras/. This low excitation gas is not taken into account in
the present model.

\begin{table*}
\caption{Observations and model predictions.
The model fluxes for each component are already
multiplied by the ISO SWS aperture correction factor (given in
parenthesis for component~1, component~2 being small enough to be
entirely seen by both ISO spectrometers) and by
the covering factor corresponding of each component. 
\label{tab:results}}
\begin{center}
    \begin{tabular}{lccrlcc}
    \hline
      \noalign{\smallskip}
    \multicolumn{1}{c}{Line} & \multicolumn{5}{c}{Line fluxes (10$^{-18}$ W/cm$^2$)} & Model/Observation \cr
         &Observations$^1$ & Model & \multicolumn{2}{r}{Component 1}&
    Component 2 & \cr
      \noalign{\smallskip}
        \hline
\allo{H}{i}{4.05}   & 11.4 $\pm$0.4    & 11.4 & .900 &(.16) & 10.5 & 1.00 \cr
\forb{Ar}{ii}{6.98} & 27.1 $\pm$3.0    & 36.3 & 1.34 &(.08) & 34.9 & 1.34 \cr
\forb{Ar}{iii}{8.98}& 20.5 $\pm$1.4    & 13.1 & 2.21 &(.23) & 10.9 & 0.64 \cr
\forb{S}{iii}{18.7} & 64.7 $\pm$1.2    & 55.5 & 25.0 &(.21) & 30.4 & 0.86  \cr
\forb{S}{iii}{33.6} & 44.3$^2$ $\pm$3.0& 23.8 & 20.8 &(.18) & 2.95 & 0.54$^2$ \cr
\forb{S}{iv}{10.5}  & 7.90 $\pm$0.8    & 8.95 & 7.70 &(.51) & 1.25 & 1.13  \cr
\forb{Ne}{ii}{12.8} &127.5 $\pm$9.0    & 97.8 & 6.08 &(.13) & 91.7 & 0.77  \cr
\forb{Ne}{iii}{15.5}& 34.0 $\pm$1.3    & 42.5 & 15.9 &(.43) & 26.6 & 1.25 \cr
\forb{Ne}{iii}{36.0}& 3.28$^2$ $\pm$1.0& 2.37 & 1.18 &(.38) & 1.18 & 0.72$^2$\cr
\forb{N}{ii}{121.8} & 4.66 $\pm$0.2    & 4.65 & 4.50 & & .150 & 1.00 \cr
\forb{N}{iii}{57.3} & 23.0 $\pm$0.7    & 23.1 & 22.6 & & .540 & 1.00 \cr
\forb{O}{iii}{51.8} & 47.4 $\pm$0.2    & 47.6 & 45.6 & & 2.08 & 1.00 \cr
\forb{O}{iii}{88.3} & 19.7 $\pm$0.4    & 19.6 & 19.4 & & .210 & 0.99 \cr
\hline
      \noalign{\smallskip}
  \multicolumn{1}{c}{Wavelength (freq.)} & \multicolumn{5}{c}{Continuum
flux density (Jy)}  &\cr
      \noalign{\smallskip}
        \hline
2~cm (15 GHz)       & 3.90 $\pm$0.5    & 3.90 & 1.34 & & 2.57 & 1.00 \cr
6~cm (5 GHz)        & 3.40 $\pm$0.2    & 3.38 & 1.53 & & 1.84 & 0.99 \cr
21~cm (1.4 GHz)     & 2.60 $\pm$0.2    & 1.88 & 1.60 &  & .282 & 0.72 \cr
        \hline
    \end{tabular}
\end{center}
$^1$ {\footnotesize The line fluxes (from \papit/) are corrected for
the interstellar extinction using A$_K=1.6$ and the law
described in \papiit/.}

$^2$ {\footnotesize Line measured with the SWS band 4 detector, see \papit/.}
\end{table*}

            \section{Discussion}
\label{sec:disc}

The detailed photoionization model of \iras/ reproduces with good
accuracy most of the atomic fine-structure line fluxes and radio flux
densities. It allows one to derive the elemental
abundances 
in the gas phase and the properties of the ionizing star(s). 
In the following, we will investigate how some of the less
constrained parameters influence the results and discuss the
reliability of the derived abundances. 

\subsection{Effect of the uncertainties of the observed line
fluxes on the model parameters}
\label{sub:errors}

The observed line fluxes are known with 10 to 20~\%
accuracy \papip/. The effect of these uncertainties on the
model are not always linear. For example, changing \ro/ by $\pm 20$~\%
changes the electron density of component~1 by $^{+34}_{-25}$~\%.
On the other hand, as the stellar effective temperature diagnostics are
extremely sensitive (as shown in Fig.~\ref{fig:t4}),
any change by $\pm 20$~\%
in any of these diagnostic line fluxes will have virtually no effect on the
determination of the stellar temperature.

Concerning the number of ionizing photons, the product of the number
of stars by the stellar luminosity is directly proportional to the
2~cm flux density.

One of the most critical parameters is the contribution of each component
to the total line fluxes. Once the density of component~2 is
determined from the 6~cm flux density, the ratio of the covering
factors is derived by fitting the 
\bra/ line flux. However, this line is sensitive to
attenuation and aperture size corrections.
As given in Table~\ref{tab:results}, the nitrogen and oxygen lines
are emitted mainly (96 to 99~\%)
by the diffuse component where only 10~\%
of the hydrogen lines emission is observed.
Decreasing the observed line flux of \bra/ by 10~\%
increases the contribution of component~1 from 36~\%
to 48~\%
and the density of component~2 from 5.2$\times 10^4$ to 9.0$\times
10^4$ cm$^{-3}$. 
The abundances of N, O, Ne, S, and Ar change by $-25, -30, +10, 0,$
and $+20$~\%
respectively.

\subsection{Filling factor and components geometry}
\label{sub:fill}

The filling factor allows one to
artificially increase the geometrical thickness of the ionized
gas. The geometry affects both the low and high ionized species if the 
thickness of the nebula is of the order of its radius.

As component~1 represents the diffuse gas, a filling factor of 1.0 seems
appropriate (the predicted extension of the gas is 35'',
i.e. comparable to the size of the observed surrounding molecular gas).
Lowering this filling factor to 0.5 has an effect on the lines mainly produced
by component~1, i.e.
\forb{N}{II}{121.8}, 
\forb{N}{III}{57.3},
\dforb{O}{III}{51.8}{88.3},
\forb{S}{iii}{33.6},
and \forb{S}{IV}{10.5}. 
The ratio \ro/ remains the same while the
\forb{N}{II}{121.8}/ \forb{N}{III}{57.3} ratio increases by about 15~\%.
A small increase of the effective temperature from 29.7 to
30.1~kK is enough to recover the observed ratio. After the whole convergence
process is performed, an increase of the N and O
abundances of about 15~\% is found.
Furthermore, as the geometrical thickness of component~1 increases up
to 45'', the effects of the finite size of the ISO SWS beam
are amplified (the \forb{S}{IV}{10.5} line significantly decreases).
No value lower than 0.5, implying greater geometrical extension, would
be acceptable.

For component~2, changing the filling factor from 1.0 to 0.5 increases the
thickness by a factor of about two. No obvious effect is found on the
line fluxes, but the radio flux densities are affected because the
self absorption is decreasing with the filling factor. The predicted 6~cm value
is then higher than
the observed value by 8~\%,
and we have to change the hydrogen inner density of component~2 from 5.2 to
9.2$\times 10^4$ cm$^{-3}$ to recover it. The geometrical thickness
of component~2 finally decreases from 1.5 to 1.0$\times 10^{16}$ cm, the
changes due to the filling factor being approximatively compensated by the
increase of density imposed by the 6~cm flux density constraint.
However, the line fluxes and the element abundances do not change
significantly.

Decreasing further the filling factor of component~2 to a value of 0.1
leads to a different behavior. The hydrogen
density needs to be increased to 5.5$\times 10^5$ cm$^{-3}$ in order to
recover the optical thicknesses of the radio
continuum at various frequencies. Such a high density implies a
collisional 
de-excitation of some
lines in component~2 (see \papiit/ for the critical densities of all
the lines). Finally, the abundances are greater than those
given in Table \ref{tab:results} by 4, 7, 77, 83, 111~\%
for N, O, Ne, S, Ar, respectively. Oxygen and nitrogen are not very
affected as these lines are emitted in component~1.
The geometrical thickness of component~2 becomes 1.6$\times
10^{15}$ cm. We could interpret this model as a distribution of very dense,
small clumps embedded in the low density medium.

In summary, modifying the filling factor leads to a change of the geometry of
the ionized gas. The main effect is on the self-absorption of the radio
free-free emission; changing the filling factor is the same as
changing the optical
depth at the different radio frequencies, in other words, changing the
proportion of the gas seen tangentially with respect to the amount of
gas seen radially.

\citet{ACHK94} derived a filling factor between 0.03 and 0.4 combining
the emission measure obtained from the continuum flux density with the 
local density obtained from the line-to-continuum ratio. They found
high values for the density (some 
10$^4$cm$^{-3}$) and they considered the gas as included in a sphere:
``assuming that the line-of-sight depth is equal to the angular
diameter from the continuum images''. In our case, the dense gas is
located in a shell (in which the filling factor is $\sim$~1.0) with a
thickness 1/20 of its radius, leading to a total filling factor for
the sphere of 0.13, compatible with the value obtained by \citet{ACHK94}.
In the model presented in this paper, the
gas is distributed in a shell, at fixed radius from the ionizing source. A more
complex model could be constructed with a distribution of clouds at
various radii, but the new free parameters introduced in such a model
could not be constrained by any available observable.

\subsection{Role of the inner radius}
\label{sub:rayon}
We fixed the inner radius of the \hii/ region to $3\times 10^{17}$ cm,
corresponding to 3'', about the radio core size
\citep[e.g.,][]{FGCV95}. This is also virtually the outer radius of the dense
component, as the geometrical thickness is 1/20 times the inner radius.
Lowering the value to e.g. $10^{17}$ cm will require
to decrease the density
of component~2 to $2.2\times 10^4$ cm$^{-3}$ in order to recover the radio
flux densities break between 2 and 6~cm.
The geometrical thickness of component~2 is now $\sim 2\times
10^{17}$ cm, still compatible with imaging observations. Oxygen and
nitrogen abundances must be decreased by 10~\%.
As the density of component~2 decreases, some lines previously
de-excited by collisions in component~2 are now emitted:
\forb{Ne}{iii}{15.5} and \forb{S}{iv}{10.5} are predicted
to be 3 and 5 times higher than the observed values.

Increasing the inner radius for component~2 will
lead to an increase of the density of this component to
recover the radio break. As the geometrical
thickness will then decrease to less than one percent
of the radius (it was 5~\%
for the adopted model, see Table~\ref{tab:model}), the self absorption
at 6~cm does not increase anymore.
This is mainly due to the spherical geometry we used for the model. A
more complicated geometry than such a thin shell could lead to more
self-absorption. We then relax the constraint of the 6~cm
flux density
and perform a model where the 6~cm predicted flux density can be
higher than the 
observation by some 10~\%.
With a density of $8\times 10^4$ cm$^{-3}$ for component~2, and
an increase or N, O, Ne, S, and Ar abundances of 12, 12, 24, 30, and 10~\%
respectively, the result is close to the adopted model.

\subsection{Role of dust}
\label{sub:dust}

The effect of adding dust in the \hii/ region is to increase the
absorption of ionizing photons and to change the shape of the
``apparent'' ionizing spectrum. We compared the emitted
spectra of the dust to the observed infrared continuum, at wavelengths 
lower than 20 $\mu$m. At longer wavelengths, the emission is dominated 
by cold dust from the PDR and the outer \hi/ region, which is not
modeled by the photoionization 
code. We check that, whatever the dust type used (i.e.
"astronomical" silicates, olivines, amorphous carbon or graphite),
the modeled emission does not exceed
the observational data for any concentration of dust lower than
$5 \times 10^{-4}$ (in mass, relative
to hydrogen). This represents an upper limit of the amount of dust, as
part of the 5-20 $\mu$m emission can be due to high temperature PAH's
present in the PDR and behind. With this amount of dust, the models have to be
adjust by changing the number of stars from 1.5 to 1.6 and the stellar 
effective temperature from 29.7 to 29.4~kK, without changes in the
abundances. In this ``maximum'' dust model, 8\% of the incoming energy
is absorbed by dust in the \hii/ region, compared to 11\% by ions.
The relatively small amount of dust derived by our modeling is in agreement
with the recent results of \citet{AE01} who found that dust in the
ionized region of S\,125 is severely depleted. A more detailed study
of the dust emission observed in \iras/, including the \hi/ region, is
postponed to a future paper.

\subsection{Number of stars versus luminosity and age}
\label{sub:nstars}

\begin{figure}
\epsfxsize=8.cm \epsfysize=7.5cm  \epsfbox{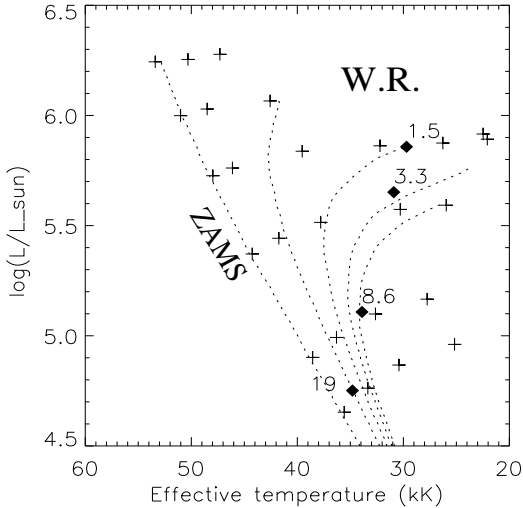}
\caption{HR diagram showing the locations of the available {\em CoStar}
models (crosses) and the four models discussed in text (filled diamonds
with corresponding number of stars). 
The dotted lines show solar metalicity isochrones for ages 0, 1.6,
2.8, 3.5 and 4.0 Myr from left to right, from the tracks of
\citet{MMS94}. \label{fig:nlt}}
\end{figure}

The number of ionizing photons is constrained by the 2~cm radio
flux density. It is a degenerated parameter since it is the  product
of the number of stars 
by the number of ionizing photons produced by one star. This
degeneracy can be explored by changing the luminosity of the
individual stars and by adjusting the corresponding value of the
number of stars.

If one changes the luminosity of the individual
stars, the effective temperature of each star must be changed in order
to recover the \forb{N}{ii}{121.7}/\forb{N}{iii}{57.3} ratio. All the
fluxes are then reproduced  as in the adopted model presented in
Table~\ref{tab:results} within a few percent.
Fig.~\ref{fig:nlt} shows in an HR diagram the locations of the
available {\em CoStar} models (crosses) and the four models 
retained and discussed here (filled diamonds). 
The number of stars needed to reproduce the 2~cm flux density is
given for each model. Although the range in effective
temperature seems small (from 30 to 35~kK), it is large if one
considers the strong constraint from the 
\forb{N}{ii}{121.7}/\forb{N}{iii}{57.3} ratio (see Fig.~\ref{fig:t4}).
Along the track between the four models with 1.5, 3.3, 8.6 and 19
stars, the stellar age varies approximatively from 2.8, 3.5, 4.0 to
1.6x$10^6$~yr.
Whatever the exact number of stars involved in the ionization of
\iras/, we see that these stars occupy a rather small range in 
effective temperature (30 to 35~kK) and age (1.6 to 4 Myr).
The obtained ages are quite old compared to ``classical'' expectations 
for \uchii/ regions. 

We cannot find a satisfying model
corresponding to one single star, as the luminosity will then overstep
the {\em CoStar} models grid and enter the post main sequence and/or 
Wolf Rayet (see
Fig. \ref{fig:nlt}). Once the temperature is derived from the
diagnostic lines ratio \forb{N}{ii}{121.7}/\forb{N}{iii}{57.3}, we
used the most luminous star available and multiply its
flux by a factor 1.5 to reproduce the radio flux
\footnote{In practice the SED we used in the model is a weighted sum 
of D4 (70~\%) and D5 (30~\%) models from \citet{SC97}, i.e. a 
$2.9\times 10^6$ years old O7 star of 50 M$_\odot$.}.
As we know from NIR observations (see \masection/~\ref{sub:stars})
that only one
star is the primary source of ionization, we think our model with 1.5
star is better. We can interpret the value of 1.5 star as a
consequence of mixing one main ionizing star with one or more lower
luminous stars.

\subsection{Dependence on atmosphere models}
\label{sub:atmomod}

The derived parameters of the ionizing source depend on the adopted
atmosphere models. Given the limited amount of constraints available
on the ionizing fluxes (cf.\ Sect.\ \ref{sub:costar}) and the potential 
uncertainties of the {\em CoStar} models especially for cool stars with 
weak winds \citep{SC97}, we have also tested other non-LTE model
atmospheres. A full description is given in \citet{MSB01}. 
Here we summarize the main effects.

We have used the recent line blanketed models of \citet{PHL01} and
test calculations for O stars based on the comoving  
frame code {\em CMFGEN} of \citet{HM98} which
both include  
stellar winds. Spectra from the fully blanketed plane parallel non-LTE
{\em TLUSTY} models of \citet{HL95} were also
kindly made available to us by Thierry Lanz.
The comparison of the predicted IR fine-structure line ratios with
observations from the sample of Paper I \& II shows that 
a consistent fit of all four ratios ([N~{\sc iii}]/[N~{\sc ii}],
[Ar~{\sc iii}]/[Ar~{\sc ii}], [S~{\sc iv}]/[S~{\sc iii}],
[Ne~{\sc iii}]/[Ne~{\sc ii}]) within a factor of two is only obtained
with the {\em CoStar} models. 
The scatter in the effective temperature
determined with the {\em CoStar} models is due to the [Ar~{\sc
iii}]/[Ar~{\sc ii}] ratio, which have a similar ionization potential
than [N~{\sc iii}] (29.6 and 27.6~eV respectively), showing a
potential problem in either the observed line fluxes, the attenuation
correction process, or in the atomic 
data. The other three excitation ratio, tracing ionizing photons
between 27.6 and 40.1~eV, are in a very good agreement.
as also seen for \iras/.

Using the extreme excitation ratio of [N~{\sc iii}]/[N~{\sc ii}] and
[Ne~{\sc iii}]/[Ne~{\sc ii}], we estimate effective temperatures of 
$\sim$ 32--35, 33--38 and 34--38 kK, using the models {\em CMFGEN} of
\citet{HM98}, {\em  
TLUSTY} of \citet{HL95}, and {\em WM-Basic} of \citet{PHL01} respectively,
while the same ratio leads to lower and less scattered effective
temperatures (29.5--30.5 kK) using the {\em CoStar} models (see
fig. \ref{fig:t4}). 

\subsection{Other constraints
and comparison with earlier studies of the exciting star}
\label{sub:otherstars}

Various estimates of the properties of the dominant ionizing star of
G29.96-0.02 have been obtained from the following observations/methods:
\begin{enumerate}
\item the bolometric luminosity $L$, estimated from IR or
   multi-wavelength observations 
\item the photon flux in the Lyman continuum $N_{\rm Lyc}$ assuming a
   single star,
\item the ratio $N_{\rm Lyc}/L$ for a stellar cluster,
\item IR line ratios through nebular modeling, 
\item the He$^+$ abundance derived from radio recombination lines,
\item near-IR photometry, and
\item K-band spectral classification of the central source.
\end{enumerate}
The estimated spectral types quoted in the literature reach from O3 
to O9.5 \citep[e.g.,][]{LBG82,DPDB94,screh95,ACW97,WCSHC97,KBHC01}.
However, these estimates are partly based on incompatible hypothesis
such as different assumptions on the source distance and different
spectral type vs.\ \teff\ calibrations. Furthermore
only unevolved zero age main sequence (ZAMS) stars were considered
in most cases, in conflict with recent evidence \citep[][this
paper]{ACW97}.
For these reasons we here briefly rediscuss these estimates using
also up-to-date stellar models and homogeneous assumptions.
In particular all observational constraints are scaled to a distance
of 6 kpc (cf.\ Sect.\ \ref{sub:dist}).
As \teff\ is the physically most important parameter -- independently
of the exact spectral-type vs.\ \teff\ relation -- we
essentially derive the constraint on this parameter.

{\bf 1)} The total bolometric luminosity of \iras/ obtained from
the 12-100 $\mu$m IRAS flux and its overall SED 
in an arcminute sized region
is $\log L/L_\odot$ $\sim$ 5.90 \citep[\papit/,][]{ACW97}.
Likely the major fraction of it is due to the 
main ionizing source \citep[cf.][and below]{ACW97}.
Using the calibrations of \citet{SK82}  yields
the following \teff : $\sim$ 44 kK (for luminosity class V), 
41 kK (LC III), 37 kK (LC I).
From \citet{VGS96} one obtains:
$\sim$ 48 kK (LC V), 45 kK (LCIII), 36 kK (LC I).
A very wide range of \teff\ ($\ga$ 20 kK to $\la$ 50 kK) is allowed 
for main sequence stars with the given luminosity, 
as shown in Fig.\ \ref{fig:nlt}.
These \teff\ represent upper limits as other stars contribute
to the total bolometric (cf.\ below).

{\bf 2)} The ionizing photon flux derived for \iras/
from radio emission is $\log(N_{\rm Lyc}) \sim$ 49.--49.14 s$^{-1}$
\citep{FGCV95,Kim01}. A somewhat higher value of 
$\log(N_{\rm Lyc}) \sim$ 49.29 was derived by \citet{ACW97} 
from the extinction corrected Brackett-$\gamma$ map.
Based on the \citet{VGS96} calibrations this corresponds
to \teff\ $\sim$ 40--43 kK (LC V), 35--38 kK (LC III), and 30--32 kK (LC I).
Similarly, the stellar models of \citet{SC97} (based on the tracks
shown in Fig.\ \ref{fig:nlt}) reproduce the observed 
$N_{\rm Lyc}$ for a temperature range between $\sim$ 30 and 46 kK, 
depending on the evolutionary state.

{\bf 3)} Given the obvious importance of small number statistics for
the number of massive stars observed in the cluster associated
with \iras/ \citep[see][]{ACW97,PMB99} standard evolutionary
synthesis models cannot be used for comparisons of $N_{\rm Lyc}/L$
\citep{CLC00}.
However, an analysis of the resolved stellar content provides further
insight.
As discussed by \citet{ACW97,PMB99} we have taken the objects 
with $H-K \ge 1.$ as cluster members.
Assuming a mean extinction $A_H=3.6$ \citep{ACW97} and using the 
synthetic photometry
of \citet{LS01} we have determined
from the H-band magnitude $m_H$ the luminosity of the individual stars 
assuming all members to be on the same isochrone with ages between 
$\sim$ 0 and 4 Myr.
From this we derive the fraction of $L$ provided by the
ionizing star, which is found to be $\sim$ 70--50~\% for ages 0--4 Myr.
A somewhat smaller fraction (70--30~\%) is found using $m_K$ 
(and $A_K=2.14$). This quantitative estimate confirms the expectations
of \citet{ACW97} of a contribution of at least 50~\% from the
ionizing star to $L$.
Correcting for $\sim$ 50 \% of $L$ due other cluster members and assuming
that one star dominates the ionization, we obtain a revised 
$N_{\rm Lyc}/L$ of the ionizing star which should be comparable to predictions
for single stars.
The comparison with the stellar models used earlier yields
\teff\ between $\sim$ 31 and 38~kK.

{\bf 4)} \citet{screh95} and \citet{ACW97} used line measurements
from KAO and photoionization models including plane parallel LTE Kurucz
model atmospheres. Their analysis (method 4) yields $T_{\rm eff}=$
35.7 and 37.5 kK respectively, rather similar to our values derived with
different atmosphere models, but larger than the value obtained with 
the {\em CoStar} atmosphere models. The main difference with their result
likely originates from the use of more sophisticated model atmospheres.

{\bf 5)} 
The He$^+$/He ionization fractions derived for the best model presented in
Table~\ref{tab:results} are 50~\% (35~\%) for Component~1 (2)
respectively, slightly lower than the values obtained by
\citet{Kim01}: 68 to 76~\%. 
Using hottest stars with {\it CMFGEN} at 33~kK and {\it
WM-Basic} at 36~kK atmosphere models, we found
He$^+$/He to be 60~\% (48\%) and 77~\% (68~\%) respectively, in better
agreement with the value obtained by \citet{Kim01} (but see the discussion
on the X$^{i+1}$/X$^{i}$ ratios for Ar, Ne, N and S in
Sect. \ref{sub:atmomod}).

{\bf 6)} From photometric observations and constraints on the total
luminosity of \iras/ \citep{ACW97} derive an allowed temperature range 
for the ionizing star of \teff\ $\sim$ 28--37 (23--43) kK at
for 1 (3) $\sigma$ uncertainties valid for source distances between
approx.\ 5--9 kpc.
Our above analysis of the cluster photometric data, taking the 
the contribution of all individual objects to $L$ into account,
yields consistency only for ages $\sim$ 3--4 Myr.
Despite this, the permitted \teff\ range based the H or K band data
remains fairly large, and essentially identical to the above values.

{\bf 7)} 
\citet{WH97} obtained the first K-band spectrum of the ionizing star of
\iras/, whose spectral type was found between O5 and O8
\citep[luminosity class undetermined; cf.][]{WH97}, based 
on the presence of He~{\sc ii} absorption, and C~{\sc iv} and N~{\sc iii}
emission lines. They note, however, that a O7 or O8 spectral type
would require some enhancement of the C~{\sc iv} and N~{\sc iii} features
-- attributed to a higher metallicity -- compared to ``normal'' objects
of these types.
While the recent VLT spectrum presented in the preliminary work of
\citet{KBHC01}  
appears to be consistent with the data of \citet{WH97}, the former authors
deduce a spectral type as early as O3 based on the presence of the C~{\sc iv} 
and N~{\sc iii} emission lines.
From this it appears that a more detailed analysis of the data of
\citet{KBHC01} 
should be awaited before more firm conclusions on the spectral type of \iras/ 
can be drawn.

In any case, given the unknown luminosity class the following 
temperature ranges are obtained for O5--O8 (O3):
$\sim$ 38.5--46 kK (51 kK) for LC V, intermediate values fo LC III, and 
$\sim$ 36--45 kK   (50 kK) for LC Ia using the \citet{VGS96} compilation
based on analysis using pure H-He atmosphere models. 
Recent fully line blanketed non-LTE calculations including stellar winds
show, however, that -- as already suspected earlier -- the \teff\ scale
of O stars must be cooler \citep[e.g.,][]{FC00,M01}.

The models of \citet{M01} yield a reduction of \teff\ by 4
to 1.5 kK  
for O3--O9.5 dwarfs compared to the \citet{VGS96} scale, and larger reductions
are expected for giants and supergiants.
Taking these effects into account we estimate for O5--O8 types \teff\ 
$\sim$ 36--43 kK for LC V and $\sim$ 33--40 kK for LC Ia.

Combining the available data it appears that
the preliminary spectral classification by \citet{KBHC01} is the only 
information which is incompatible with most other constraints (points 3-6,
possibly also 1 and 2). 
Good consistency is obtained, however, from the intersection of the above 
constraints 1) to 6), yielding an allowed \teff\ between $\sim$ 31 and 37 kK, 
overlapping with the spectral type derived by \citet{WH97}. 
We thus conclude that overall the parameters derived from our photoionization 
modeling are compatible with all the available observational data.

\begin{table*}
\caption{Abundances determined by previous authors and in the present
work.\label{tab:compar}}
\begin{center}
    \begin{tabular}{lccccccccc}
    \hline
    Element
    & \citet{HHF81}
    &\multicolumn{2}{c}{\citet{screh95}}
    &\multicolumn{2}{c}{\citet{ACW97}}
    &\multicolumn{2}{c}{\papiit/}
    & This work
    & Solar$^2$
    \cr
    & \iras/
    & \iras/
    & 4.5 kpc$^1$
    & \iras/
    & 4.5 kpc$^1$
    & \iras/
    & 4.5 kpc$^1$
    &
    &
    \cr
    \hline
    N/H (10$^{-4}$)
    & --
    & 2.3
    & 1.8
    & 1.8
    & 1.2
    & 1.9
    & --
    & 2.0
    & 0.8
    \cr
    O/H (10$^{-4}$)
    & --
    & 8.5
    & 6.6
    & 5.6
    & 7.3
    & 5.1
    & --
    & 4.6
    & 6.8
    \cr
    Ne/H (10$^{-4}$)
    & 2.7
    & 2.6
    & 1.8
    & --
    & --
    & 2.5
    & 2.2 
    & 1.7
    & 1.2
    \cr
    S/H (10$^{-5}$)
    & 3.2
    & 1.9
    & 1.6
    & 2.2
    & 1.8
    & 0.8 
    & --
    & 2.2
    & 2.1
    \cr
    Ar/H (10$^{-6}$)
    & 23.
    & --
    & --
    & --
    & --
    & 4.8
    & 5.0 
    & 5.0
    & 2.5
    \cr
    \hline
     N/O
    & --
    & 0.27
    & 0.27
    & 0.32
    & 0.17
    & 0.37
    & 0.33
    & 0.43 
    & 0.12
    \cr
    Ne/S
    & 8.4
    & 13.
    & 11.
    & --
    & --
    & 36. 
    & -- 
    & 7.5
    & 5.7
    \cr
        \hline
    \end{tabular}
\end{center}
$^1$ {\footnotesize Values obtained applying the gradients from the
corresponding authors at the galactocentric distance of \iras/ }

$^2$ {\footnotesize from \citet{Solar}.} 

\end{table*}

\subsection{Implications of the advanced age of \iras/}
\label{sub:comp_star}

However, an age of $\simeq 3\times 10^6$ years for the star is very
high compared to 
the expected dynamical lifetime of \uchii/ regions \citep[$5\times
10^5$ years, 
see e.g.,][based on the number of \uchii/ regions in the Galaxy and
their expected lifetime]{WC89b}.
Two main models have been developed to
explain the cometary morphology which is
common for \uchii/ regions. Models of stellar-wind bow shocks
\citep[see e.g.,][]{MLVBWC91}, due to an O star moving supersonically
through a molecular cloud, were first studied and applied to \iras/
\citep{WC91,VBML92,LH96}. Champagne flow models
\citep[see e.g.,][]{YTTB83}, resulting from the expansion of
an \hii/ region into a molecular cloud exhibiting a density gradient, are
also able to reproduce the cometary morphology. These models were applied
more recently to \iras/ \citep[][ PMB99]{FGCV95,LH96} and
were found to give results more consistent with the radio observations.

It is important to note that assuming a reasonable projected proper
motion of 1~km/s, the star 
should have moved away from its birth place by about 3~pc (1.75~arcmin) in
$\sim 3\times 10^6$ years. This rules out the Champagne flow model as
a complete description of \iras/ and
strongly favors the random meeting of an older star with an interstellar
cloud. The ionizing star may also have left its birthplace,
irradiating molecular gas further out which could
still be part of the larger parental cloud from 
which it was formed. 

\subsection{Reliability of the abundances determination}

The determination of
the elemental abundances from the infrared fine-structure lines
depends on many physical parameters, such as the filling factor, which
are poorly constrained. Nevertheless,we can assert that there
are two groups of 
elements. On one hand, oxygen and nitrogen, whose lines, all observed
by the LWS spectrometer, are mostly emitted by the extended component~1,
due to their low critical densities. Uncertainties in the attenuation
correction and then in the \bra/ line flux by, e.g., 10~\%
leads to an uncertainty on the N and O abundances of 25 to 30~\%
(see \masection/~\ref{sub:errors}).

The elements neon, argon and sulfur group, whose lines
are observed by the SWS spectrometer (as the \hi/ lines) with all
the subsequent aperture corrections, are emitted by both components.
The presence of high density clumps
(filling factor of 0.1 -- see \masection/~\ref{sub:fill}) in the
core will lead to abundances two times higher than what we
determined in the presented model for the Ne, Ar, S group. 

Whatever the
uncertainties could be on the filling factor, the geometry of the
source, the attenuation or the actual value of the radio emission, the
determination of the abundance ratios in each group are robust: the
N/O ratio on one hand, and the Ne/Ar, Ne/S and Ar/S ratios on the other.

Table~\ref{tab:compar} compares the abundances determined here
and by \citet{HHF81}, \citet{screh95}, \citet{ACW97} and
\papiit/. The solar abundances from \citet{Solar} are also given.
\citet{ACW97} used the \citet{screh95} observations
to modeled \iras/, but with a core/halo
description. They both made semi-empirical models (using icf's) and
adopted an higher effective temperature ($\sim$~36~kK,
see discussion in \masection/~\ref{sub:otherstars}). The method
used in \papiit/ is semi-empirical, based on the same observed line fluxes
as the present work.
For those previous works, we give the values effectively determined for
\iras/ and the values obtained using the abundance gradient law they
found, applied at the position of \iras/: 4.5~kpc from the galactic
center. 

The set of abundances, exepted for oxygen, shows that \iras/ is
overabundant compared to the solar values, in agreement with its
inner position in the Galaxy. 
The abundances determined in the present work are compatible
with the previous determination within a factor of 2, except with the
Ar/H ratio from \citet{HHF81} and S/H ratio from \papiit/.

The determination of the sulfur abundance relative to hydrogen in
\papiit/ is very different from what the previous authors and the
present work found. From the results presented in
Table~\ref{tab:results} we see that the \forb{S}{iii}{18.7} and
\forb{S}{iv}{10.5} lines on which the sulfur abundance is based in
\papiit/ are mostly emitted by the extended component~1. The
effect of the finite aperture size of the SWS instrument is crucial in
this case. As there was no correction for this effect in
\papiit/, the sulfur emission and its abundance are underestimated.

            \section{Conclusion}
\label{sec:concl}

This paper presents a detailed model of the compact Galactic
\hii/ region \iras/ for which high quality data imagery and 
spectroscopy is available at both infrared and radio wavelengths,
including recent ISO observations. The model, which is based on the
photoionization code NEBU and state-of-the-art stellar atmosphere
models, reproduces most of the observations, with the exception
of a few points known to be less accurately measured. The radio
and infrared data on \iras/ are best reproduced by a 2-density 
component model nebula, with a diffuse (n$_H\sim 600$ cm$^{-3}$)
extended (1~pc) halo surrounding a dense (n$_H\sim
50000$ cm$^{-3}$) compact (0.1~pc) core.

Using {\em CoStar} stellar atmosphere models we derived 
an effective temperature of $\sim 30^{+2}_{-1}$ kK.
Adopting more recent non-LTE line blanketed atmospheres with
stellar winds, a somewhat higher \teff\ $\sim$ 32--38 kK is found.
This temperature range is compatible with all observational
constraints. For \teff\ $\sim$ 33-36 kK compatibility is also obtained 
with the K-band spectral type O5--O8 determined by \citep{WH97}
when recent downward revisions of the effective temperature 
scale of O stars \citep{M01} are taken into account.

We explored the effect of varying the different model parameters 
on the predictions. The main sources of uncertainty in determining
the abundances are the fluxes of the hydrogen recombination lines 
and the geometry of the dense compact core.

The derived elemental abundances are in agreement with the lowest
values found in previous studies. The most robust results are N/O and 
Ne/S which are 3.5 and 1.3 times the solar values, respectively.

From the reanalysis of the different available observational constraints
(see Sect.\ \ref{sub:otherstars}) it is not surprising that several earlier 
studies reached apparently conflicting results on the spectral type
or \teff\ of the ionizing source of \iras/.
This is mostly due to the following facts.
First, proper photoionization models must account for the dependence
of nebular lines on several parameters, including the ionization parameter,
geometry etc.
Second, consistent predictions from stellar models regarding ionizing
fluxes, \teff\ scales etc. should be used taking into account recent
progress made with fully line blanketed non-LTE atmosphere models.
Last, but not least, as the available K-band spectroscopy does not
allow a determination of the luminosity class, allowance should be
made for possible variations of the evolutionary state of ionizing
sources in compact \hii/ regions.
In view of these issues one may question whether mid-IR analysis
of compact \hii/ regions yield intrinsically different estimates 
of \teff\ compared to spectral types or other constraints
as suggested by \citet{HLR02}. Our detailed analysis of \iras/
indicates the opposite. 

The age of the ionizing star required by our model is 
$\approx 3 \times 10^6$~yr, much older than the expected lifetime of \uchii/
regions. This could indicate that \iras/
is not excited by a bona fide young massive star. Instead
the ionizing star creating today the \hii/ region \iras/ 
may have left its birthplace, exciting gas further out. 
This matter could still be part of the larger parental 
cloud from which the stellar cluster associated with
\iras/ was formed.


\acknowledgements

We thank the referee for useful questions and comments.
CM thanks D. P\'equignot for helpful discussions on
photoionization models and Ryszard Szczerba for discussions on dust.
DS thanks Alan Watson, Margaret Hanson, and Yuri Izotov for 
useful discussions and Margaret Hanson for sharing data before 
publication.
We thank Thierry Lanz for sending us line blanketed {\em TLUSTY}
model atmospheres before publication, and John Hillier for
making his atmosphere code {\em CMFGEN} available.



\end{document}